\documentclass[twocolumn,aps,prd,superscriptaddress,showpacs,nofootinbib,floatfix]{revtex4-1}
\usepackage{epsfig,dcolumn,bm,amsmath}

\newcommand{\dmu}{\partial_\mu}

\newcommand{\lm}{{\cal M}}
\newcommand{\scL}{{\cal L}}

\newcommand{\oh}{\frac{1}{2}}
\newcommand{\trh}{\frac{3}{2}}

\newcommand{\sqd}{\sqrt{2}}
\newcommand{\sqt}{\sqrt{3}}
\newcommand{\sqs}{\sqrt{6}}
\newcommand{\sqdt}{\sqrt{\frac{2}{3}}}
\newcommand{\sqtd}{\sqrt{\frac{3}{2}}}
\newcommand{\squt}{\frac{1}{\sqt}}
\newcommand{\squd}{\frac{1}{\sqd}}
\newcommand{\squs}{\frac{1}{\sqs}}
\newcommand{\be}{\begin{eqnarray}}
\newcommand{\ee}{\end{eqnarray}}
\newcommand{\ba}{\begin{array}}
\newcommand{\ea}{\end{array}}
\newcommand{\bt}{\begin{tabular}}
\newcommand{\et}{\end{tabular}}
\newcommand{\btab}{\begin{table}}
\newcommand{\etab}{\end{table}}
\newcommand{\bc}{\begin{center}}
\newcommand{\ec}{\end{center}}
\newcommand{\nn}{\nonumber}

\newcommand{\Eq}[1]{Eq.~(\ref{#1})}

\newcommand{\coma}{\textrm{, }}
\newcommand{\scB}{{\cal B}}

\begin{document}

\title{Radiative decays of dynamically generated charmed baryons}

\author{D. Gamermann}{\thanks{E-mail: daniel.gamermann@ific.uv.es}}
\author{C. E. Jim\'enez-Tejero}
\author{A. Ramos}
\affiliation{Departament d'Estructura i Constituents de la Mat\`eria and Institut de Ci\`encies del Cosmos,
Universitat de Barcelona, Avda. Diagonal 647, E-08028 Barcelona, Spain}

\begin{abstract}
In this work we study the radiative decay of dynamically generated
$J^P=\oh^-$ charm baryons into the ground state $J^P=\oh^+$
baryons. Since different theoretical interpretations of these baryonic resonances, and in particular of the $\Lambda_c(2595)$,
give different predictions, a precise experimental measurement of these decays would be an important step for understanding their nature.
\end{abstract}


\maketitle


\section{Introduction}

The research on charm hadron physics has experienced a renewed interest in the last years, after the discovery of a few meson states which could not be easily described as $q{\bar q}$ states of quark model potentials, such as the $D_{s0}(2317)$, the $D_{s1}(2460)$ \cite{Aubert:2003fg,Briere:2006ff,Krokovny:2003zq,Abe:2003jk}, in the open-charm sector, and the $X(3872)$ \cite{Choi:2003ue,Acosta:2003zx,Abazov:2004kp,Aubert:2004ns}, the $X(3940)$ \cite{Abe:2007jn,Abe:2007sya}, the $Y(3940)$ \cite{Abe:2004zs,Aubert:2007vj} or the $Z(3930)$ \cite{Uehara:2005qd}, in the hidden-charm sector. There has also been a substantial amount of theoretical progress leading to the observation that some of these states could be interpreted alternatively as molecules \cite{Barnes:2003dj,Kolomeitsev:2003ac,Hofmann:2003je,Guo:2006fu,Guo:2006rp,Zhang:2006ix,Close:2003sg,Wong:2003xk,Swanson:2003tb,Gamermann:2006nm,Gamermann:2007fi,Gamermann:2009fv}, tetraquarks \cite{Chen:2004dy,Maiani:2004vq,Bracco:2005kt,Bicudo:2005de}, or hybrids \cite{Li:2004sta}. A similar situation is found in the baryon sector. Some of the newly observed charmed baryons states   \cite{2880-Artuso:2000xy,Mizuk:2004yu,Jessop:1998wt,Csorna:2000hw,Chistov:2006zj,Aubert:2007dt,Aubert:2006je,2940-Aubert:2006sp,2880-Abe:2006rz}, such as the $J^P=\oh^-$ $\Lambda_c(2595)$ or the $J^P=\trh^-$ $\Lambda_c(2625)$, can be interpreted as meson-baryon molecules that can be generated dynamically within coupled-channel unitary schemes \cite{Tolos:2004yg,Lutz:2003jw,lutz5,Hofmann:2005sw,Hofmann:2006qx,Mizutani:2006vq,GarciaRecio:2008dp,JimenezTejero:2009vq,meunegc}.

The decay modes of a resonance may provide a way of learning about the nature of the state: whether it fits in the conventional $qqq$-baryon or $q\bar q$-meson picture or it has a more exotic interpretation. In particular, the radiative decays into lower lying states may represent a significant part of the decay width when the hadronic modes are suppressed by phase-space restrictions or/and small values of the coupling constants. Note that electromagnetic transitions are also useful in the determination of the quantum numbers of states decaying into a final hadron with well established quantum numbers.

Radiative decays of ground state heavy flavored baryons have been studied within Heavy Hadron Chiral Perturbation Theory (HHCPT) \cite{Cheng:1992xi,Cheng:1997rp,Savage:1994wa,Banuls:1999br} --an approach which combines Heavy Quark (HQ) symmetry with the chiral symmetry in the light sector--, employing light-cone QCD sum-rules \cite{Zhu:1998ih}, supplementing Heavy Quark symmetry with a SU$(2N_f)\times$O(3) symmetry in the light diquark system \cite{Tawfiq:1999cf}, or implementing also dynamical effects for the internal quark structure of the hadrons
within the relativistic three-quark model (RTQM) \cite{Ivanov:1999bk,Ivanov:1998wj} or other non-relativistic quark models \cite{Dey:1994qi,Fayyazuddin:1997jn,Majethiya:2009vx}.

Many of the former approaches also have been applied to obtain the radiative decays of excited heavy flavored baryons, such as the $\Lambda_c(2595)$. A first qualitative estimation was given in Ref.~\cite{Cho:1994vg}, where the
HHCPT formalism was extended to include the lowest lying excited baryon doublet, $\Lambda_c(2595)$ ($J^P=\oh^-$) and $\Lambda_c(2625)$ ($J^P=\trh^-$). The results of Ref.~\cite{Tawfiq:1999cf} were obtained by exploiting the alternative light diquark SU$(2N_f)\times$O(3) symmetry. Radiative decays of excited charmed baryons were also calculated within the relativistic quark model of Refs.~\cite{Ivanov:1999bk,Ivanov:1998wj}, as well as using light-cone QCD sum-rules in Ref.~\cite{Zhu:2000py}. All these works consider the excited heavy baryon as being an orbital excitation of the three-quark $Qqq$ system with a unit of angular momentum inserted between the heavy quark $Q$ and the light diquark $qq$. A different perspective is provided by the model of Ref.~\cite{Chow:1995nw}, which considers the excited $\Lambda_c(2595)$ and $\Lambda_c(2625)$ as being $D^*N$ bound systems N in the first excited state of a harmonic potential adjusted phenomenologically to reproduce the spin-averaged excitation energy, or by the model of Ref.~\cite{Dong:2010xv} which considers the radiative decay of the $\Lambda_c(2940)$ in a hadronic $D^* N$ molecular picture.



In the present work we study the radiative decays of excited baryons in the $C=1$ sector. We focus on the resonances $\Lambda_c(2595)$, $\Sigma_c(2800)$, $\Xi_c(2790)$ and $\Xi_c(2970)$, all having $J^P=\oh^-$, which have been generated as meson-baryon molecular states on several models based on coupled-channel dynamics \cite{Tolos:2004yg,Lutz:2003jw,lutz5,Hofmann:2005sw,Hofmann:2006qx,Mizutani:2006vq,GarciaRecio:2008dp,JimenezTejero:2009vq}. The radiative transitions to baryons of the  $J^P=\oh^+$  ground state multiplet proceeds, as we will see, through the coupling of the photon to the various meson-baryon components of the resonance, as determined by the coupled-channel dynamical model of  Ref.~\cite{JimenezTejero:2009vq}. This is essentially different from the quark models for which, in the heavy quark limit, the photon only couples to the light diquark system. Therefore, a precise measurement of the radiative decays of excited heavy flavored baryons would help in understanding their internal structure.

The radiative decays of dynamically generated charm mesons has already been addressed recently. In \cite{meurad,fasrad,lutz1}, the radiative decay of the $D^*_{s0}(2317)$ meson has been studied from the point of view that it is generated dynamically mainly from the interaction of the $D$ meson with a kaon. Also radiative decays of the puzzling $X(3872)$ has been calculated from the point of view that this state is a $D^*\bar{D}+c.c.$ molecule \cite{yubid}. More recently, many different radiative decays of the controversial $X$, $Y$ and $Z$ states have been analyzed assuming that their structure is determined by the interactions of two vector mesons \cite{molina1,molina2,molina3}

This work is organized as follows: in the next section we explain the model for generating dynamically the $J^P=\oh^-$
baryon resonances from the interaction of ground state baryons with pseudo-scalar mesons. In Sec. \ref{sec:rad} we
present the framework for evaluating the diagrams needed in the calculation of the radiative decays of the resonances.
The results of the calculation are presented and discussed in Sec. \ref{sec:res} and in Sec. \ref{sec:conc} we present
a brief overview and our conclusions.


\section{The dynamically generated states}
\label{sec:states}

In this section we will review the coupled-channel approach employed in \cite{JimenezTejero:2009vq} to obtain dynamically generated open-charm baryon resonances.

The scattering amplitudes $T$, which describe the scattering of the pseudo-scalar meson fields off the ground-state baryon fields
can be obtained by solving the well-known Lipmann--Schwinger equation, which schematically reads
\be
T=V+VJT\ .
\label{eq:ls}
\ee
The loop function $J$ is the product of the meson and baryon single-particle propagators, and the scattering kernel $V$ describes
the interaction between the pseudo-scalar mesons and the ground-state baryons. Following the original work of Hofmann and Lutz \cite{Hofmann:2005sw},
we identify a $t$-channel exchange of vector mesons as the driving force for the $s$-wave scattering between pseudo-scalar mesons in 16-plet and baryons
in 20-plet representations. The scattering kernel takes the form (see \cite{Hofmann:2005sw} for details)
\begin{align}
\begin{split}
V^{(I,S,C)}_{ij}(p_i,q_i,p_j,q_j)=\frac{g^2}{4}\sum_{V \in [16]} C^{(I,S,C)}_{ij;V}\bar u(p_j)\gamma^\mu \\
\left(g_{\mu\nu}-\frac{(q_i-q_j)_\mu (q_i-q_j)_\nu}{m^2_V}\right)
\frac{1}{t-m^2_V}(q_i+q_j)^\nu u(p_i) \ ,
\label{eq:sk}
\end{split}
\end{align}
where the sum runs over all vector mesons of the SU(4) $16$-plet, $(\rho$, $K^*$, $\bar K^*$, $\omega$, $\phi$,
$D^*$, $D_s^*$, $\bar D^*$, $\bar D_s^*$, $J/\Psi)$, $m_V$ is the mass of the exchanged vector meson, $g$ is the universal vector meson coupling constant, $p_i, q_i$ ($p_j,q_j$) are the four momenta of the incoming (outgoing) baryon and meson, and the coefficients $C^{(I,S,C)}_{ij;V}$ denote the strength of the interaction
in the different isospin ($I$), strangeness ($S$), charm ($C$) sectors, and meson-baryon channels $(i,j)$. For the coupling constant $g$ we use the value $6.6$, which reproduces the decay width of the
$\rho$ meson \cite{pdg}. The $s$-wave projection of the scattering kernel is easily obtained and, in the center-of-mass (c.m.)
frame, it takes the analytical form
\begin{align}
\begin{split}
V_{ij,l=0}^{(I,S,C)}(\vec k_i,\vec k_j)=
N\frac{g^2}{8}
\sum_{V \in[16]} C^{(I,S,C)}_{ij;V} \\
\left[ \frac{2\beta}{b} +\frac{\alpha b-\beta a}{b^{2}}   \ln\left(\frac{a+b}{a-b}\right)\right] \ ,
\label{eq:t}
\end{split}
\end{align}
with $a, b, \alpha$ and $\beta$ being
\be
a&=&m^2_i+m^2_j-2\omega_i(|\vec k_i|) \omega_j(|\vec k_j|)-m^2_V  \, \nonumber \\
b&=&2 |\vec k_i| |\vec k_j| \, \nonumber \\
\alpha &=& \Omega_i(|\vec k_i|)+\Omega_j(|\vec k_j|)-M_i-M_j  \nonumber \\
&&-\frac{m^2_j-m^2_i}{m^2_V}(\Omega_j(|\vec k_j|)-\Omega_i(|\vec k_i|)+M_i-M_j) \, \nonumber \\
\beta&=&\frac{|\vec k_i||\vec k_j| }{(E_i(|\vec k_i|)+M_i)(E_j(|\vec k_j|)+M_j)}  \nonumber \\
&&\left(\Omega_i(|\vec k_i|)+\Omega_j(|\vec k_j|)+M_i+M_j  \right. \nonumber \\
&&\left. -\frac{m^2_j-m^2_i}{m^2_V}(\Omega_j(|\vec k_j|)-\Omega_i(|\vec k_i|)+M_j-M_i)\right) \ ,
\ee
where $\vec k_i,\vec k_j$ are the initial and final relative momenta, $m_i$, $m_j$, $M_i$, $M_j$ are the masses
of the incoming and outgoing mesons and baryons, $\omega_i(|\vec k_i|)$, $\omega_j(|\vec k_j|)$, $E_i(|\vec k_i|)
$, $E_j(|\vec k_j|)$ their corresponding energies, which have been taken to be their
on-shell values, and $\Omega(|\vec k|)$ stands for the total energy $\omega(|\vec k|)+ E(|\vec k|)$ of the meson-baryon pair.  The factor $N=[(E(|\vec k_i|)+M_i)(E(|\vec k_j|)+M_j)/(4M_i M_j)]^{1/2}$
comes from the normalization of the Dirac spinors.


Once the scattering kernel has been constructed, it can be inserted in the s-wave projected Lipmann--Schwinger equation,
\begin{align}
\begin{split}
T^{(I,S,C)}_{ij,l=0}(\vec k_i,\vec k_j,\sqrt{s})=V^{(I,S,C)}_{ij,l=0}(\vec k_i,\vec k_j)+\sum_{m}\int\frac{d\vec k}{(2\pi)^3}\\
F(|\vec k|)V^{(I,S,C)}_{im,l=0}(\vec k_i,\vec k)J_m(\sqrt{s},\vec k)T^{(I,S,C)}_{mj,l=0}(\vec k,\vec k_j,\sqrt{s}) \,
\label{eq:lspw}
\end{split}
\end{align}
where the loop function $J$ explicitly reads
\begin{align}
\begin{split}
J^{(I,S,C)}_m(\sqrt{s},\vec k)=\frac{M_m}{2E_{m}(|\vec k|)\omega_{m}(|\vec k|)} \\
\frac{1}{\sqrt{s}-E_{m}(|\vec k|)-\omega_{m}(|\vec k|)+i\eta} \ ,
\label{eq:j}
\end{split}
\end{align}
and a dipole-type form factor $F(|\vec k|)$,
\be
F(|\vec k|)=\left(\frac{\Lambda^2}{\Lambda^2+|\vec k|^2}\right)^2 \ ,
\label{eq:ff}
\ee
have been introduced in order to regularize the integral. This form is typically adopted in studies of hadron-hadron interactions within the scheme
of Lipmann-Schwinger-type equations in the light flavor sector \cite{machleidt}. The value of the cut-off $\Lambda$ is a free parameter of our model that is fitted to the existing data. In Table \ref{tab:summary} we collect,
from the resonances generated dynamically in
Ref.~\cite{JimenezTejero:2009vq} in various $C=1$ sectors, those states that, using a cut-off
value of around 1 GeV, can be readily identified with observed resonances \cite{pdg}.

\bc
\begin{table}[htbp]
\caption{Masses and widths of states that can be
identified with well established resonances in various sectors of $C=1$.}\label{tab:summary}
\begin{tabular*}{0.48\textwidth}{@{\extracolsep{\fill}} c|c|cc|cc}
\hline
\hline
$(I,S)$ & $\Lambda$ & \multicolumn{2}{c|}{Theory} &  \multicolumn{2}{c}{Experiment}\\
 & [MeV] & $M_R$ & $\Gamma$ & $M_R$& $\Gamma$ \\
 & & [MeV] & [MeV] & [MeV] & [MeV] \\
\hline
\hline
$(0,0)$ & 903 & 2595  & 0.5 & $2595.4 \pm 0.6$ & $3.6^{+2.0}_{1.3}$ \\
$\Lambda_c$ &   &  & &  & \\
\hline
$(1,0)$ & & &  &  $2801^{+ 4}_{-6}$ & $75^{+22}_{-17} ~~(\Sigma_c^{++})$  \\
$\Sigma_c$ & 1100& 2792&16 & $2792^{+ 14}_{- 5}$ & $62^{+60}_{-40}~~(\Sigma_c^+)$ \\
 & & & &   $2802^{+4}_{- 7}$ & $61^{+28}_{-18}~~(\Sigma_c^0)$ \\
 \hline
$(1/2,-1)$ & 980 & 2790 & 0.5 & $2791.8 \pm 3.3$ & $<12 ~~(\Xi_c^0)$ \\
$\Xi_c$ &  & \multicolumn{2}{c|}{} &\multicolumn{2}{c}{}\\
 & 960 & 2970 & 5.1 & $2968.0 \pm 2.6$ & $20\pm 7~~(\Xi_c^0)$ \\
\hline
\hline
\end{tabular*}
\end{table}
\ec

In order to associate a given enhancement of the scattering amplitude to a
resonance, we look for a characteristic pole, $z_R$, in the unphysical sheet of the
complex energy plane. 
Our prescription of unphysical sheet is such that,
whenever the real part of the complex energy crosses a meson-baryon threshold
cut, the sign of the on-shell momentum is changed for this channel and for
the already opened
ones, as described in detail in Ref.~\cite{Logan:1967zz}.
If the pole  $z_R$ lies not too far from the real axis, its value determines
the Breit-Wigner mass ($M={\rm Re\,}z_R$) and width ($\Gamma=2\,{\rm
Im\,}z_R$) of the resonance as seen from real energies. 
The couplings of the resonance into its various meson-baryon components
are obtained from the residues of the scattering amplitude since, close to the
pole, it can be parameterized in the form:
\be
T_{ij,l=0}^{I,S,C}(\vec k_i,\vec k_j,z)=\frac{g_ig_j}{z-z_R} \ .
\label{eq:pole}
\ee
Note that the value of the coupling constants of
Eq.~(\ref{eq:pole}) depends on the particular momentum values chosen in the
evaluation of the $T$-matrix element. However, this dependence is very mild and we have obtained the couplings from the amplitudes at $\vec k_i=\vec k_j=0$. The values of the couplings of the resonances to the various charged meson-baryon states are listed in Tables \ref{tab:g_RBM1} to \ref{tab:g_RBM3}.

\bc
\begin{table}[htbp]
\caption{Coupling coefficients $g_{RBM}$ of the $\Lambda_c(2595)$ pole to the different channels} \label{tab:g_RBM1}
\bt{c|c}
\hline
\hline
&$g_{RBM}$\\
$MB$& $2595 +i 0.25$ [MeV]\\
\hline
&  \\
 ${\pi\Sigma_{c}}$ &$0.06-i  0.31$\\
 ${DN}$ & $-0.13+i  11$\\
 ${\eta\Lambda_{c}}$&$-0.005+i  0.42$\\
 ${K\Xi_{c}}$  &$-0.006+i  0.52$\\
 ${K\Xi'_{c}}$ &$0.03-i  0.23$\\
 ${D_{s}\Lambda}$ &$-0.06+i  6$\\
 ${\eta'\Lambda_{c}}$ &$-0.01+i  0.69$\\
 ${\eta_{c}\Lambda_{c}}$ &$0.02-i  2$\\
 ${\bar{D}\Xi_{cc}}$ &$0.007+i  0.05$\\
&  \\
\hline
\hline
\et
\end{table}
\ec

\bc
\begin{table}[htbp]
\caption{Coupling coefficients $g_{RBM}$ of the $\Sigma_c(2792)$ pole to the different channels} \label{tab:g_RBM2}
\bt{c|c}
\hline
\hline
&$g_{RBM}$ \\
$MB$& $2792 +i  8.16$ [MeV]\\
\hline
&   \\
 $\pi \Lambda_{c}$ &$0.19+i  0.53$\\
 $ \pi \Sigma_{c}$ & $-0.28+i  0.12$\\
 $DN $ &$-0.60-i  3.6$\\
$K\Xi_{c} $ &$0.31+i  0.28$\\
$\eta\Sigma_{c} $ &$0.06-i  0.21$\\
$K\Xi'_{c} $ & $-0.13+i  0.24$\\
$ D_{s}\Sigma$ &$0.42+i  3.1$\\
$ \eta'\Sigma_{c}$ &$-0.05-i  0.35$\\
$\bar{D}\Xi_{cc} $ &$0.22+i  0.28$\\
$\eta_{c}\Sigma_{c}$&$0.13+i  1$\\
&   \\
\hline
\hline
\et
\etab
\ec
\bc
\begin{table}[htbp]
\caption{Coupling coefficients $g_{RBM}$ of the $\Xi_c(2790)$ and $\Xi_c(2970)$ poles to the different channels} \label{tab:g_RBM3}
\bt{c|c|c}
\hline
\hline
& \multicolumn{2}{c}{$g_{RBM}$} \\
$MB$&$2790+i 0.25$ [MeV] & $2970 +i 2.5$ [MeV]\\
\hline
& &  \\
 $\pi \Xi_{c}$   &$0.01+i0.06$ &$0.008-i0.37$  \\
 $\pi \Xi'_{c}$   &$0+i0.002$ &$-0.25-i0.08$  \\
 $\bar{K}\Lambda_{c}$  &$-0.06-i0.22$ &$0.1+i0.17$  \\
 $ \bar{K}\Sigma_{c}$ &$-0.003-i1.5$ &$-0.17+i0.03$ \\
 $D\Lambda$  &$0.08-i2.1$ &$-0.21-i3.2$  \\
 $\eta\Xi_{c} $ &$-0.07-i0.33$ &$0.08-i0.02$  \\
 $D\Sigma $  &$-0.03+i12.$ &$-0.07-i1.76$  \\
 $\eta\Xi'_{c}$  &$-0.001-i0.73$ &$-0.02-i0.05$  \\
 $K\Omega_{c} $ &$-0.003+i0.08$ &$-0.21+i0.15$ \\
 $ D_{s}\Xi$  &$-0.08+i6.54$ &$0.18+i2.8$ \\
 $\eta'\Xi_{c}$ &$0.005-i0.76$ &$-0.007-i0.08$ \\
 $\eta'\Xi'_{c}$ &$0.004+i0.16$ &$-0.01-i0.31$ \\
 $\bar{D}_{s}\Xi_{cc}$  &$0.04+i0.44$ &$-0.02-i0.006$ \\
 $\bar{D}\Omega_{cc} $  &$-0.03+i0.03$ &$-0.03+i0.04$ \\
 $\eta_{c}\Xi_{c}$  &$-0.007-i1$ &$0.05+i0.91$ \\
 $\eta_{c}\Xi'_{c}$ &$-0.01-i0.46$ &$0.04+i0.9$  \\
 & &  \\
\hline
\hline
\et
\etab
\ec

\section{Radiative decay calculation} \label{sec:rad}

The reaction we study is given by:
\be
B^*(P,\chi_i)&\rightarrow&\gamma(K,\epsilon) B(Q,\chi_f)
\ee
where $P$, $K$ and $Q$ are four-momenta which are related by $P=K+Q$, $\epsilon$ is the photon polarization and $\chi_i$ and $\chi_f$
are the polarization of the initial and final spin $\oh$ baryons $B^*$ and $B$, respectively.

\begin{figure}
\begin{center}
\includegraphics[width=7cm]{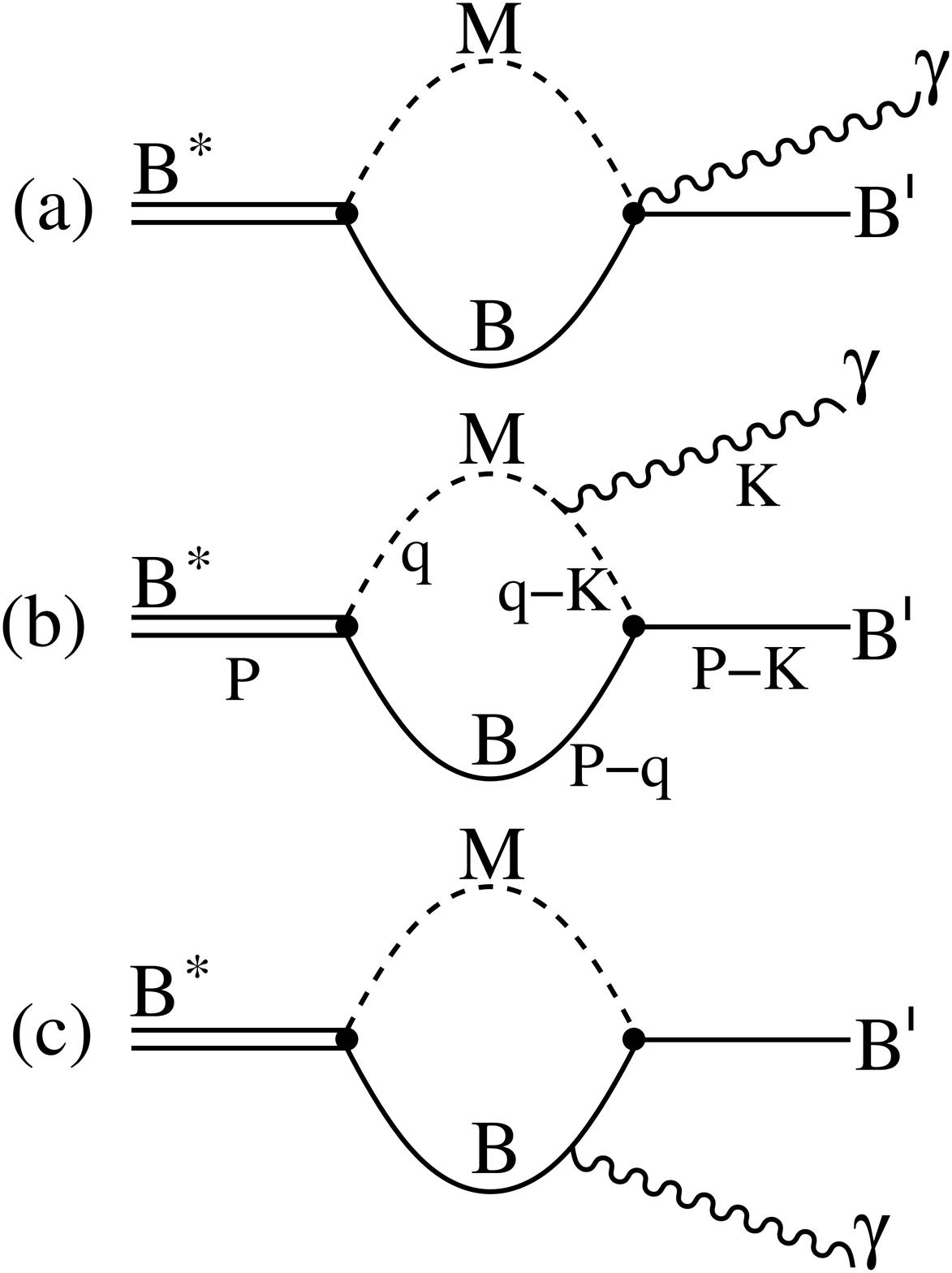}
\caption{Diagrams needed for the evaluation of the radiative decay of dynamically generated baryons.} \label{fig:diags}
\end{center}
\end{figure}

For the evaluation of these radiative decays, we follow a gauge invariant scheme already used for non-charmed resonances
\cite{meisner,doering1,doering2,geng}. As shown there, the mechanisms for the decay of the resonance $B^*$ into $\gamma B$ are given by the diagrams of 
 Fig.~\ref{fig:diags}, where the photon couples to the constituent mesons and baryons. The amplitude for the radiative decay has the structure:
\be
-i\lm&=&T_{\mu\nu}\epsilon^\nu\bar{\chi_f}\sigma^\mu\chi_i
\label{eq:amplitude}
\ee
with $\sigma_\mu=(0,\vec{\sigma})$, where $\sigma_i$ are the usual Pauli matrices. For the tensor $T_{\mu\nu}$, there are
five possible independent Lorentz structures that one can construct with the two independent four-momenta:
\be
T^{\mu\nu}&=&a g^{\mu\nu}+b P^\mu P^\nu+c P^\mu K^\nu\nn\\
 &+&d K^\mu P^\nu+e K^\mu K^\nu \label{eq:abcde}
\ee
This expression is simplified by noting that, due to the Lorentz condition $\epsilon_\nu K^\nu=0$, the terms with the coefficients $c$ and $e$ in \Eq{eq:abcde} will not contribute to the radiative decay amplitude of Eq.~(\ref{eq:amplitude}). Moreover, gauge invariance imposes
$T_{\mu\nu} K^\nu=0$ and we obtain:
\be
\big(a+d(P.K)\big) K^\mu+b(P.K)P^\mu&=&0 \ ,
\ee
from where we conclude that $b=0$, while
$a$ and $d$ are related through $a=-d(P.K)$. Finally, working
in the rest frame of the decaying baryon ($\vec{P}=\vec{0}$) and
taking the Coulomb gauge ($\epsilon^0=0$), only the $a$ term contributes to the amplitude. The $a$ coefficient is more easily calculated through
the $d$ coefficient, using the above mentioned relation. This is so
because, as we will show, the integral for evaluating the $d$
coefficient converges and it receives contributions from only the (b) and (c) diagrams of
Fig.~\ref{fig:diags}.



The amplitude for diagram \ref{fig:diags}(b) is given by:
\be
-i\lm&=&-ig_{RBM}g_{BBM}g_{\gamma MM}\int\frac{d^4q}{(2\pi)^4}\frac{1}{(q-K)^2-m_M^2}\nn\\
&\times&\frac{1}{q^2-m_M^2}\frac{2m_B}{(P-q)^2
-m_B^2}\bar{\chi}_f(q-K).\sigma\chi_i\nn\\
&\times&(2q-K).\epsilon
\ee
where $g_{RBM}$ is the coupling of the resonance to its meson and baryon constituents, $g_{BBM}$ is the coupling of the
baryonic current to a meson, $g_{\gamma MM}$ is the coupling of the meson to a photon, and $m_M$ and $m_B$ are masses of the
meson and baryon in the loop, respectively. Using the Feynman parameterization
\be
\frac{1}{abc}&=&2\int_0^1dx\int_0^{1-x}dy\nn\\
&\times&\frac{1}{\big(x(a-b)+y(c-b)+b\big)^3} \ ,
\ee
and the identity
\be
\int d^4q^\prime\frac{1}{(q^{\prime 2}+s+i\epsilon)^3} = \frac{i\pi^2}{2}\frac{1}{s+i\epsilon} \ ,
\ee
together with the relation between the $a$ and $d$ terms of the amplitude, one finds the following contribution of diagram \ref{fig:diags}(b) to the $a$ term:  \be
&&-i\lm^a_{\rm{(b)}}=g_{RBM}g_{BBM}g_{\gamma MM}\frac{2m_B}{(4\pi)^2}\int_0^1dx\int_0^{1-x}dy\nn\\
&& \phantom{-i\lm^a_{\rm{(b)}}} \times\frac{x(y-1)}{s+i\epsilon}2P.K \label{eq:diagb} \\
&&\ \ \ s=x\big(P^2(1\!-\!x)\!-\!m_B^2\!+\!m_M^2\!-\!2yP.K\big)\!-\!m_M^2
\ee

Analogously, one can calculate the contribution coming from the diagram in Fig. \ref{fig:diags} (c) as:
\be
&&-i\lm^a_{\rm{(c)}}=g_{RBM}g_{BBM}g_{\gamma BB}\frac{2m_B}{(4\pi)^2}\int_0^1dx\int_0^{1-x}dy\nn\\
&& \phantom{-i\lm^a_{\rm{(c)}}} \times \frac{-yx}{s+i\epsilon}2P.K \label{eq:diagc} \\
&&\ \ \ s=x\big(P^2(1\!-\!x)\!-\!m_M^2\!+\!m_B^2\!-\!2yP.K\big)\!-\!m_B^2
\ee

With respect to the couplings appearing in \Eq{eq:diagb} and \Eq{eq:diagc}, the values of $g_{RBM}$, relating the resonance
to its constituents, are listed in Tables  \ref{tab:g_RBM1} to \ref{tab:g_RBM3} of previous section, where the channels appearing there are given in the isospin basis.
The couplings of the mesons
to the photon, $g_{\gamma MM}$, and those of the baryons to the photon, $g_{\gamma BB}$, are $e Q$ where $Q$ is the electric charge $(-1,0,1$ or $2)$ of the meson or the
baryon which emits the photon. Finally,
the coupling of the mesons to the ground state baryons is given by:
\be
g_{BBM}&=&\alpha_i\frac{D+F}{2f}+\beta_i\frac{D-F}{2f}  \label{eq:coupbbm}
\ee
where $D+F$=1.26, $D-F$=0.33 \cite{ramos1}. We have taken
$f=1.15f_M$ with $f_M=f_\pi=93$ MeV for light mesons and
$f_M=f_D=165$ MeV
for heavy mesons. The coefficients $\alpha_i$ and $\beta_i$, which depend on the SU(4) flavor structure of the interaction, have been obtained
from the Lagrangian for the interaction of the ground state
spin $\oh^+$ baryons and the pseudo-scalar mesons, which is constructed in the following.

Our prescription of
field definitions for the baryons and mesons is that of Ref.~\cite{Hofmann:2005sw}. The baryons are represented by a three index tensor $\scB^{ijk}$, antisymmetric in the first two indices, where $i$, $j$ and $k$ run from 1 to
4:
\be
\scB^{121}&=&p\coma \scB^{122}=n\coma \scB^{132}=\squd\Sigma^0-\squs\Lambda\coma \nn\\
\scB^{124}&=&\frac{2}{\sqs}\Lambda_c\coma \scB^{141}=-\Sigma_c^{++}\coma \scB^{142}=\squd\Sigma_c^++\squs\Lambda_c\coma\nn\\
\scB^{143}&=&\squd\Xi_c^{\prime+}-\squs\Xi_c^+\coma \scB^{144}=\Xi_{cc}^{++}\coma \scB^{213}=\frac{2}{sqs}\Lambda\coma\nn\\
\scB^{231}&=&\squd\Sigma^0+\squs\Lambda\coma \scB^{232}=\Sigma^-\coma \scB^{233}=\Xi^-\coma\nn\\
\scB^{234}&=&\frac{2}{\sqs}\Xi_c^0\coma \scB^{241}=\squd\Sigma_c^+-\squs\Lambda_c\coma \scB^{242}=\Sigma_c^0\coma\nn\\
\scB^{243}&=&\squd\Xi_c^{\prime 0}+\squs\Xi_c^0\coma \scB^{244}=-\Xi_{cc}^+\coma \scB^{311}=\Sigma^+\coma\nn\\
\scB^{313}&=&\Xi^0\coma \scB^{314}=\frac{2}{sqs}\Xi_c^+\coma \scB^{341}=\squd\Xi_c^{\prime+}+\squs\Xi_c^+\coma\nn\\
\scB^{342}&=&\squd\Xi_c^{\prime0}-\squs\Xi_c^0\coma \scB^{343}=\Omega_c\coma \scB^{344}=\Omega_{cc}
\ee
The pseudo-scalar meson field is represented by a matrix:
\begin{widetext}
\be
\Phi&=&\left( \begin{array}{cccc}
 \frac{\pi^0}{\sqd}+\frac{\eta_8 }{\sqs}+\frac{\eta_{15}}{\sqrt{12}} & \pi^+ & K^+ & \bar D^0 \\ & & & \\
 \pi^- & \frac{\pi^0}{\sqd}+\frac{\eta_8 }{\sqs}+\frac{\eta_{15}}{\sqrt{12}} & K^0 & -D^- \\& & & \\
 K^- & \bar K^0 & \frac{-2\eta_8}{\sqs}+\frac{\eta_{15}}{\sqrt{12}} & D_s^- \\& & & \\
 D^0 & -D^+ & D_s^+ & \frac{-3\eta_{15}}{\sqrt{12}} \\ \end{array} \right)
\ee
\end{widetext}

The Lagrangian for the $BBM$ interaction reads:
\be
\scL_{BBM}&=&\frac{\sqd}{4f}\sum_{i,j,k,l=1}^4\bar{\scB}_{ijk}\gamma^\mu\bigg((D+F)\dmu\Phi_{kl}\scB_{ijl}\nn\\
&-&2(D-F)\dmu\Phi_{jl}\scB_{ilk}\bigg)  \label{eq:lag} \ .
\ee

Note that the physical $\eta$, $\eta^\prime$
and $\eta_c$ fields are related to the mathematical $\eta_8$ and $\eta_{15}$ fields, belonging to the pseudo-scalar meson fifteen-plet, and the singlet $\eta_1$ field. The mixture is given by:
\be
\left(\ba{c} \eta_1 \\ \\
             \eta_8 \\ \\
             \eta_{15}\ea\right) &=& \left(\ba{ccc} \frac{1}{2\sqt}&\sqdt&\oh \\ & & \\
                                                    \frac{2\sqd}{3}&-\frac{1}{3}&0 \\ & & \\
                                                     \frac{1}{6}&\frac{\sqd}{3}&-\frac{\sqt}{2}\ea\right).\left(\ba{c} \eta \\ \\
                                                     \eta^\prime \\ \\
                                                     \eta_c\ea\right) \ .
\ee
Therefore, in order to obtain a Lagrangian for the physical fields, we we need to add to \Eq{eq:lag}  an interaction of the
baryon current with the singlet field $\eta_1$:
\be
&& \scL^\prime_{BBM} = \scL_{BBM} \nn \\
&&\ \ +\frac{\sqd}{4f}(D\delta_D
+F\delta_F)\oh\sum_{i,j,k=1}^4\bar{\scB}_{ijk}\gamma^\mu\scB_{ijk}\dmu \eta_1 \ , \label{eq:lagfin}
\ee
where the coefficients $\delta_D=-\squd$ and $\delta_F=\frac{3}{\sqd}$ are obtained by consistently imposing that processes where the $\eta_c$ meson couples to light baryons ($N$, $\Lambda$, $\Sigma$ or $\Xi$) should vanish.

The complete Lagrangian $\scL^\prime_{BBM}$ of \Eq{eq:lagfin} allows us to determine all coupling constants $g_{BBM}$, which we write in the form of Eq.~(\ref{eq:coupbbm}). The specific values of the $\alpha_i$ and $\beta_i$ coefficients for all the transitions needed in the different radiative decays are collected in Tables \ref{tab:lcsc} to \ref{tab:scpp} of the appendix.

With all the required ingredients having been established, one can finally obtain the the radiative decay width from:
\be
\Gamma&=&\frac{1}{\pi}\left|\sum_{i} {\cal M}_{i}^a \right|^2E_\gamma\frac{m_B}{m_{B^*}}
\ee
where $E_\gamma$ is the photon energy, ${\cal M}_i^a={\cal M}^a_{(b)}+ {\cal M}^a_{(c)}$, and the sum runs over all contributing intermediate $MB$ channels.


\section{Results} \label{sec:res}

We evaluate the following radiative decays:
\be
\Lambda_c(2595)&\rightarrow&\Lambda_c\gamma \\
\Lambda_c(2595)&\rightarrow&\Sigma_c^+\gamma \\
\Sigma^{+}_c(2792)&\rightarrow&\Lambda_c\gamma \\
\Sigma^{++}_c(2792)&\rightarrow&\Sigma^{++}_c\gamma \\
\Sigma^{+}_c(2792)&\rightarrow&\Sigma^{+}_c\gamma \\
\Sigma^{0}_c(2792)&\rightarrow&\Sigma^{0}_c\gamma \\
\Xi^+_c(2790)&\rightarrow&\Xi^+_c\gamma \\
\Xi^{0}_c(2790)&\rightarrow&\Xi^{0}_c\gamma \\
\Xi^+_c(2790)&\rightarrow&\Xi^{\prime+}_c\gamma \\
\Xi^0_c(2790)&\rightarrow&\Xi^{\prime0}_c\gamma \\
\Xi^+_c(2970)&\rightarrow&\Xi^+_c\gamma \\
\Xi^0_c(2970)&\rightarrow&\Xi^0_c\gamma \\
\Xi^+_c(2970)&\rightarrow&\Xi^{\prime+}_c\gamma \\
\Xi^0_c(2970)&\rightarrow&\Xi^{\prime0}_c\gamma \ ,
\ee
for $\Lambda_c(2595)$, $\Sigma_c(2792)$, $\Xi_c(2790)$ and $\Xi_c(2970)$, which are the resonances that, according to the models of Ref.~\cite{Tolos:2004yg,Lutz:2003jw,lutz5,Hofmann:2005sw,Hofmann:2006qx,Mizutani:2006vq,GarciaRecio:2008dp,JimenezTejero:2009vq}, can plausibly be interpreted as dynamically generated meson-baryon molecules.

We start showing the results for the $\Lambda_c(2595)$, which is the analog in the charm sector of the $\Lambda(1405)$ in the strange sector and, therefore, is a good candidate for being a meson-baryon molecular state. The radiative decay widths are:
\be
\Gamma_{\Lambda_c(2595)\rightarrow\Lambda_c\gamma}&=&278\textrm{ KeV} \\
\Gamma_{\Lambda_c(2595)\rightarrow\Sigma_c^+\gamma}&=&2\textrm{ KeV} \ .
\ee
These values are also collected in Table \ref{tab:reslc}, where the contribution to the width coming from each intermediate channel is also shown, together with the sign, $(+)$ or $(-)$, of the corresponding amplitude. This allows one to analyze the constructive or destructive character of the interferences between the various channels. First of all, we note the tremendous difference in size, of two orders of magnitude, between the radiative decay rate of the $\Lambda_c(2595)$ into $\Lambda_c \gamma$ and $\Sigma_c^+ \gamma$ states. To understand the origin of this difference we focus on the most important contribution, which corresponds to the $D^+ n$ intermediate state in both cases. 
First of all, the ratio between the couplings $D^+ n \to \Lambda_c$ and
$D^+ n \to \Sigma_c^+$ is 3.8, as can be inferred from the $\alpha$ and $\beta$  coefficients of Table~\ref{tab:lcsc}. Moreover, the different kinematical variables of the two processes produce a $D^+ n$ loop which is, for $\Lambda_c \gamma$ decay, a factor 2 larger than for $\Sigma_c^+ \gamma$ decay.
The square of this two factors, together with the ratio of photon energies, which are $E_\gamma=290$ MeV for $\Lambda_c \gamma$ and
$E_\gamma=138$ MeV for $\Sigma_c^+ \gamma$, explain the factor 100 difference between the $D^+n$ contributions to both decays. We also note that there are constructive interferences between the most important contributions ($D^0 p$, $D^+n$, $D^+_s\Lambda$) to $\Lambda_c \gamma$ decay, which enhance even further this rate compared to the $\Sigma_c^+ \gamma$ one.

\bc
\btab[htbp]
\caption{Results for the radiative decay of the $\Lambda_c(2595)$ compared with other theoretical approaches.
The sign in brackets indicates the sign of the amplitude, so one can know when the interference between the channels
is constructive or destructive.} \label{tab:reslc}
\bt{c|cccc|ccc}
\hline
\hline
Channel & $\Gamma_{\Lambda_c\gamma}$ & \cite{Ivanov:1999bk} & \cite{Zhu:2000py}  & \cite{Chow:1995nw} & $\Gamma_{\Sigma_c^+\gamma}$ & \cite{Ivanov:1999bk} & \cite{Zhu:2000py}\\
        & [KeV] & [KeV] & [KeV] & [KeV] & [KeV] & [KeV] & [KeV] \\
\hline
\hline
$\pi^+\Sigma_c^0$ & 1.2$(+)$ & & &   & 0.3$(-)$ & & \\
$\pi^0\Sigma_c^+$ & $<$0.1$(+)$ & & &   & 0 & & \\
$\pi^-\Sigma_c^{++}$ & 0.7$(-)$ & & &   & 0.2$(-)$ & & \\
$D^0p$ & 9.1(+) & & &   & 0.1$(+)$ & & \\
$D^+n$ & 83.4(+) & & &   & 0.8$(-)$ & & \\
$\eta\Lambda_c$ & $<$0.1$(+)$ & & &   & 0 & & \\
$K^+\Xi_c^0$ & $<$0.1$(+)$ & & &   & $<$0.1$(+)$ & & \\
$K^0\Xi_c^+$ & $<$0.1$(+)$   & & &   & $<$0.1$(-)$ & & \\
$K^+\Xi_c^{\prime0}$ & $<$0.1$(+)$ & & &   &$<$0.1$(-)$ & & \\
$K^0\Xi_c^{\prime+}$ & $<$0.1$(+)$ & & &   &$<$0.1$(+)$ & & \\
$D_s^+\Lambda$ & 13.5$(+)$ & & &   &0 & & \\
$\eta^\prime\Lambda_c$ & $<$0.1$(+)$   & & &   &0 & & \\
$\eta_c\Lambda_c$ & $<$0.1$(+)$ & & &   &0 & & \\
$\bar{D}^0\Xi_{cc}^+$ & $<$0.1$(+)$ & & &   &$<$0.1$(-)$ & & \\
$D^-\Xi_{cc}^{++}$ & $<$0.1$(-)$ & & &   &$<$0.1$(-)$ & & \\
\hline
Total & 278 & 115 & 36 &  16 & 2 & 77 & 11\\
\hline
\hline
\et
\etab
\ec

In Table~\ref{tab:reslc} we also compare our results to those obtained by other calculations performed within the relativistic quark model \cite{Ivanov:1999bk}, using light-cone QCD sum-rules \cite{Zhu:2000py}, or adopting a bound $D^* N$ picture for the $\Lambda_c(2595)$ \cite{Chow:1995nw}. We observe a large diversity of results. Note also that the
two orders of magnitude difference between the radiative decays into $\Lambda_c \gamma$ or $\Sigma_c \gamma$ states found in the present work is not obtained by any of the other models displayed in Table~\ref{tab:reslc}, nor by the HHCPT results of Ref.~\cite{Cho:1994vg}, which estimated partial widths of the same order of magnitude. 
Obviously, the tremendous differences between models calls for a measurement of these decay modes which could bring essential information about the nature of the $\Lambda_c(2595)$.

The radiative decays of the other resonances, for which there is no experimental observation nor other theoretical predictions, are:
\be
\Gamma_{\Sigma^{++}_c(2792)\rightarrow\Sigma^{++}_c\gamma}&=&51\textrm{ KeV} \\
\Gamma_{\Sigma^{+}_c(2792)\rightarrow\Sigma^{+}_c\gamma }&=&29\textrm{ KeV}\\
\Gamma_{\Sigma^{0}_c(2792)\rightarrow\Sigma^{0}_c\gamma}&=&9\textrm{ KeV} \\
\Gamma_{\Sigma_c^{+}(2792)\rightarrow\Lambda_c\gamma}&=&35\textrm{ KeV} \\
\Gamma_{\Xi^+_c(2790)\rightarrow\Xi^+_c\gamma}&=&246\textrm{ KeV} \\
\Gamma_{\Xi^{0}_c(2790)\rightarrow\Xi^{0}_c\gamma}&=&117\textrm{ KeV} \\
\Gamma_{\Xi^+_c(2790)\rightarrow\Xi^{\prime+}_c\gamma}&=&1\textrm{ KeV} \\
\Gamma_{\Xi^0_c(2790)\rightarrow\Xi^{\prime0}_c\gamma}&=&1\textrm{ KeV} \\
\Gamma_{\Xi^+_c(2970)\rightarrow\Xi^+_c\gamma}&=&55\textrm{ KeV} \\
\Gamma_{\Xi^0_c(2970)\rightarrow\Xi^0_c\gamma}&=&2\textrm{ KeV} \\
\Gamma_{\Xi^+_c(2970)\rightarrow\Xi^{\prime+}_c\gamma}&=&40\textrm{ KeV} \\
\Gamma_{\Xi^0_c(2970)\rightarrow\Xi^{\prime0}_c\gamma}&=&9\textrm{ KeV}
\ee
It would also be interesting to see whether our predictions, based in a molecular picture for these resonances, are also very different from those obtained in quark-model theoretical approaches.


\section{Conclusions and Overview} \label{sec:conc}

We have studied the radiative decay width of charmed baryon resonances, which have been obtained as poles in the $T$-matrix of a coupled channels dynamical model and, therefore, they are interpreted as composite meson-baryon molecular states.

We have obtained a sizable value for the one-photon radiative width of the $\Lambda_c(2595)$. This resonance decays radiatively mostly into $\Lambda_c \gamma$ states, with a partial rate of $\Gamma_{\Lambda_c(2595)\rightarrow\Lambda_c\gamma}=278\textrm{ KeV}$, while only a tiny amount of the width, $\Gamma_{\Lambda_c(2595)\rightarrow\Sigma_c^+\gamma}=2\textrm{ KeV}$, is due to the decay into $\Sigma_c \gamma$ final states.
Our results are very different, in size and distribution among decay channels, to what is found by other approaches in the literature.

We have also presented predictions for the radiative decay of other excited charmed baryons. The radiative decay widths of the $\Sigma_c(2800)$ and $\Xi_c(2970)$ resonances are found to be relatively small, of the order of a few tenths of KeV. However, the transitions 
$\Xi_c(2790)^+ \to \Xi_c^+ \gamma$ and $\Xi_c(2790)^0 \to \Xi_c^0 \gamma$ are substantially larger, worth exploring experimentally.

The sizable value of some widths, especially those of the $\Lambda(2595)$ and the $\Xi_c(2790)$ resonances, makes the study of these radiative reactions a very useful tool to obtain information about the characteristics of these charmed baryons. 


\section*{Acknowledgments}

The authors would like thank Isaak Vida\~na for useful discussions.
This work is partly supported by the EU contract No. MRTN-CT-2006-035482
(FLAVIAnet), by the contract FIS2008-01661 from MICINN
(Spain), by the Ge\-ne\-ra\-li\-tat de Catalunya contract 2009SGR-1289,
and by FEDER/FCT (Portugal) under the
project CERN/FP/83505/2008. We
acknowledge the support of the European Community-Research Infrastructure
Integrating Activity ``Study of Strongly Interacting Matter'' (HadronPhysics2,
Grant Agreement n. 227431) under the Seventh Framework Programme of EU.

\appendix

\section{Baryon-baryon-meson coupling constants \mbox{\boldmath{$g_{BBM}$}}}
\label{app:coup}

This Appendix gives the values of the coefficients $\alpha_i$ and $\beta_i$ that define, according to Eq.~(\ref{eq:coupbbm}), the $BBM$ coupling constants $g_{BBM}$, needed in the diagrams that determine the various radiative decays.

\bc
\btab[htb]
\caption{The $\alpha$ and $\beta$ coefficients for the channels involved in the radiative decay of resonances into
$\Lambda_c$ and $\Sigma_c^+$} \label{tab:lcsc}
\bt{c|cccc}
\hline
\hline
 & \multicolumn{2}{c}{$\Lambda_c$} & \multicolumn{2}{c}{$\Sigma_c^+$} \\
$MB$&$\alpha$&$\beta$&$\alpha$&$\beta$\\
\hline
& & & & \\
$\pi^+\Sigma_c^0$ & $-\squt$ & $-\squt$ & 1 & -1 \\
& & & & \\
$\pi^0\Sigma_c^+$ & $-\squt$ & $-\squt$ & 0 & 0 \\
& & & & \\
$\pi^-\Sigma_c^{++}$ & $-\squt$ & $-\squt$ & -1 & 1 \\
& & & & \\
$D^0p$ & $\frac{2}{\sqt}$ & $-\squt$ &0&1\\
& & & & \\
$D^+n$ & $-\frac{2}{\sqt}$ & $\squt$ &0&1\\
& & & & \\
$\eta\Lambda_c$ & $\frac{\sqd}{3\sqt}$ & $\frac{5\sqd}{3\sqt}$ &0&0\\
& & & & \\
$K^+\Xi_c^0$ & $-\frac{1}{3\sqd}$ & $\frac{5}{3\sqd}$ & $\squs$& $\squs$\\
& & & & \\
$K^0\Xi_c^+$ & $-\frac{1}{3\sqd}$ & $\frac{5}{3\sqd}$& $-\squs$& $-\squs$\\
& & & & \\
$K^+\Xi_c^{\prime0}$ & $-\squs$ & $-\squs$& $\squd$& $-\squd$\\
& & & & \\
$K^0\Xi_c^{\prime+}$ & $\squs$ & $\squs$& $\squd$& $-\squd$\\
& & & & \\
$D_s^+\Lambda$ & $-\frac{2\sqd}{3}$ & $\frac{\sqd}{3}$ &0&0\\
& & & & \\
$\eta^\prime\Lambda_c$ & $\frac{1}{3\sqt}$ & $-\frac{5}{3\sqt}$ &0&0\\
& & & & \\
$\eta_c\Lambda_c$ & $\frac{2\sqd}{3}$ & $-\frac{\sqd}{3}$ &0&0\\
& & & & \\
$\bar{D}^0\Xi_{cc}^+$ & $\squt$ & $-\frac{2}{\sqt}$ & -1&0\\
& & & & \\
$D^-\Xi_{cc}^{++}$ & $-\squt$ & $\frac{2}{\sqt}$ & -1&0 \\
& & & & \\
$\pi^0\Lambda_c$ & 0 & 0 & $-\squt$ & $-\squt$ \\
& & & & \\
$\eta\Sigma_c^+$ & 0 & 0 & $\sqdt$ & $-\sqdt$ \\
& & & & \\
$D_s^+\Sigma^0$ & 0 & 0 & 0 & $-\sqd$ \\
& & & & \\
$\eta^\prime\Sigma_c^+$ & 0 & 0 & $\squt$ & $-\squt$ \\
& & & & \\
$\eta_c\Sigma_c^+$ & 0 & 0 & 0 & $-\sqd$ \\
\hline
\hline
\et
\etab
\ec

\bc
\btab[htb]
\caption{The $\alpha$ and $\beta$ coefficients for the channels involved in the radiative decay of resonances into
$\Xi^+_c$ and $\Xi^{\prime+}_c$} \label{tab:cscp}
\bt{c|cccc}
\hline
\hline
 & \multicolumn{2}{c}{$\Xi^+_c$} & \multicolumn{2}{c}{$\Xi^{\prime+}_c$} \\
$MB$&$\alpha$&$\beta$&$\alpha$&$\beta$\\
\hline
& & & & \\
$\pi^+\Xi_c^0$ & $-\frac{1}{3\sqd}  $ & $ \frac{5}{3\sqd} $ & $ -\squs $ & $ -\squs $ \\
& & & & \\
$\pi^0\Xi_c^+$ & $ \frac{1}{6} $ & $ -\frac{5}{6} $ & $ \frac{1}{2\sqt} $ & $ \frac{1}{2\sqt} $ \\
& & & & \\
$\pi^+\Xi_c^{\prime0}$ & $ \squs $ & $ \squs $ & $ \squd $ & $ -\squd $ \\
& & & & \\
$\pi^0\Xi_c^{\prime+}$ & $ \frac{1}{2\sqt} $ & $ \frac{1}{2\sqt} $ & $ \oh $ & $ -\oh $ \\
& & & & \\
$\bar{K}^0\Lambda_c$ & $\frac{1}{3\sqd}  $ & $ -\frac{5}{3\sqd} $ & $ \squs $ & $ \squs $ \\
& & & & \\
$\bar{K}^0\Sigma_c^+$ & $ -\squs $ & $ -\squs $ & $ \squd $ & $ -\squd $ \\
& & & & \\
$K^-\Sigma_c^{++}$ & $ \squt $ & $ \squt $ & $ -1 $ & $ 1 $ \\
& & & & \\
$D^+\Lambda$ & $ -\frac{\sqd}{3} $ & $ \frac{1}{3\sqd} $ & $ 0 $ & $ \sqtd $ \\
& & & & \\
$\eta\Xi_c^+$ & $ 0 $ & $ 0 $ & $ \frac{\sqd}{3} $ & $ \frac{\sqd}{3} $ \\
& & & & \\
$D^+\Sigma^0$ & $ \sqdt $ & $ -\squs $ & $ 0 $ & $ -\squd $ \\
& & & & \\
$D^0\Sigma^+$ & $ \frac{2}{\sqt} $ & $ -\squt $ & $ 0 $ & $ -1 $ \\
& & & & \\
$\eta\Xi_c^{\prime+}$ & $ \frac{\sqd}{3} $ & $ \frac{\sqd}{3} $ & $ 0 $ & $ 0 $ \\
& & & & \\
$K^+\Omega_c$ & $ \squt $ & $ \squt $ & $ 1 $ & $ -1 $ \\
& & & & \\
$D_s^+\Xi^0$ & $ \frac{2}{\sqt} $ & $ -\squt $ & $ 0 $ & $ 1 $ \\
& & & & \\
$\eta^\prime\Xi_c^+$ & $\frac{1}{2\sqt}  $ & $ \frac{5}{2\sqt} $ & $ -\frac{1}{6} $ & $ -\frac{1}{6} $ \\
& & & & \\
$\eta^\prime\Xi_c^{\prime+}$ & $ -\frac{1}{6} $ & $ -\frac{1}{6} $ & $ \frac{\sqt}{2} $ & $ -\frac{\sqt}{2} $ \\
& & & & \\
$D_s^-\Xi_{cc}^{++}$ & $ -\squt $ & $ \frac{2}{\sqt} $ & $ 1 $ & $ 0 $ \\
& & & & \\
$\bar{D}^0\Omega_{cc}$ & $ \squt $ & $ -\frac{2}{\sqt} $ & $ 1 $ & $ 0 $ \\
& & & & \\
$\eta_c\Xi_c^+$ & $ \frac{2\sqd}{3} $ & $ -\frac{\sqd}{3} $ & $ 0 $ & $ 0 $ \\
& & & & \\
$\eta_c\Xi_c^{\prime+}$ & $ 0 $ & $ 0 $ & $ 0 $ & $ -\sqd $ \\
\hline
\hline
\et
\etab
\ec

\bc
\btab[htb]
\caption{The $\alpha$ and $\beta$ coefficients for the channels involved in the radiative decay of resonances into
$\Xi^0_c$ and $\Xi^{\prime0}_c$. We only show the coefficients for channels with charged particles.} \label{tab:cscz}
\bt{c|cccc}
\hline
\hline
 &  \multicolumn{2}{c}{$\Xi^0_c$} &  \multicolumn{2}{c}{$\Xi^{\prime0}_c$} \\
$MB$&$\alpha$&$\beta$&$\alpha$&$\beta$\\
\hline
& & & & \\
$\pi^-\Xi_c^+$ & $ -\frac{1}{3\sqd} $ & $ \frac{5}{3\sqd} $ & $ \squs $ & $ \squs $ \\
& & & & \\
$\pi^-\Xi_c^{\prime+}$ & $ -\squs $ & $ -\squs $ & $ \squd $ & $ -\squd $ \\
& & & & \\
$K^-\Lambda_c$ & $ -\frac{1}{3\sqd} $ & $ \frac{5}{3\sqd} $ & $ -\squs $ & $ -\squs $ \\
& & & & \\
$K^-\Sigma_c^+$ & $ \squs $ & $ \squs $ & $ \squd $ & $ -\squd $ \\
& & & & \\
$D^+\Sigma^-$ & $ -\frac{2}{\sqt} $ & $ \squt $ & $ 0 $ & $ -1 $ \\
& & & & \\
$D_s^+\Xi^-$ & $ \frac{2}{\sqt} $ & $ -\squt $ & $ 0 $ & $ -1 $ \\
& & & & \\
$D_s^-\Xi_{cc}^+$ & $ -\squt $ & $ \frac{2}{\sqt} $ & $ -1 $ & $ 0 $ \\
& & & & \\
$D^-\Omega_{cc}$ & $ \squt $ & $ -\frac{2}{\sqt} $ & $ -1 $ & $ 0 $ \\
\hline
\hline
\et
\etab
\ec

\bc
\btab[htb]
\caption{The $\alpha$ and $\beta$ coefficients for the channels involved in the radiative decay of resonances into
$\Sigma_c^0$. We only show the coefficients for channels with charged particles.} \label{tab:scz}
\bt{c|cc}
\hline
\hline
 & \multicolumn{2}{c}{$\Sigma^0_c$} \\
$MB$&$\alpha$&$\beta$\\
\hline
& &\\
$\pi^-\Lambda_c$ & $ -\squt $ & $ -\squt $ \\
& &  \\
$\pi^-\Sigma_c^+$ & $ 1 $ & $ -1 $ \\
& &  \\
$D_s^+\Sigma^-$ & $ 0 $ & $ -\sqd $ \\
& &  \\
$D^-\Xi_{cc}^+$ & $ \sqd $ & $ 0 $ \\
\hline
\hline
\et
\etab
\ec

\bc
\btab[htb]
\caption{The $\alpha$ and $\beta$ coefficients for the channels involved in the radiative decay of resonances into
$\Sigma_c^{++}$. We only show the coefficients for channels with charged particles.} \label{tab:scpp}
\bt{c|cc}
\hline
\hline
 &  \multicolumn{2}{c}{$\Sigma^{++}_c$} \\
$MB$&$\alpha$&$\beta$\\
\hline
& & \\
$\pi^+\Lambda_c$ & $ -\squt $ & $ -\squt $ \\
& & \\
$\pi^+\Sigma_c^+$ & $ -1 $ & $ 1 $ \\
& & \\
$\pi^0\Sigma_c^{++}$ & $ 1 $ & $ -1 $ \\
& & \\
$D^+p$ & $ 0 $ & $ -\sqd $ \\
& & \\
$K^+\Xi_c^+$ & $ \squt $ & $ \squt $ \\
& & \\
$\eta\Sigma_c^{++}$ & $ \sqdt $ & $ -\sqdt $ \\
& & \\
$K^+\Xi_c^{\prime+}$ & $ -1 $ & $ 1 $ \\
& & \\
$D_s^+\Sigma^+$ & $ 0 $ & $ -\sqd $ \\
& & \\
$\eta^\prime\Sigma_c^{++}$ & $ \squt $ & $ -\squt $ \\
& & \\
$\bar{D}^0\Xi_{cc}^{++}$ & $ -\sqd $ & $ 0 $ \\
& & \\
$\eta_c\Sigma_c^{++}$ & $ 0 $ & $ -\sqd $ \\
\hline
\hline
\et
\etab
\ec



\begin{thebibliography}{99}


\bibitem{Aubert:2003fg}
  B.~Aubert {\it et al.}  [BABAR Collaboration],
  Phys.\ Rev.\ Lett.\  {\bf 90} (2003) 242001
  [arXiv:hep-ex/0304021].

\bibitem{Briere:2006ff}
  R.~A.~Briere {\it et al.}  [CLEO Collaboration],
  Phys.\ Rev.\  D {\bf 74} (2006) 031106
  [arXiv:hep-ex/0605070].

\bibitem{Krokovny:2003zq}
  P.~Krokovny {\it et al.}  [Belle Collaboration],
  Phys.\ Rev.\ Lett.\  {\bf 91} (2003) 262002
  [arXiv:hep-ex/0308019].

\bibitem{Abe:2003jk}
  K.~Abe {\it et al.}  [Belle Collaboration],
  Phys.\ Rev.\ Lett.\  {\bf 92} (2004) 012002
  [arXiv:hep-ex/0307052].

\bibitem{Choi:2003ue}
  S.~K.~Choi {\it et al.}  [Belle Collaboration],
  Phys.\ Rev.\ Lett.\  {\bf 91} (2003) 262001
  [arXiv:hep-ex/0309032].

\bibitem{Acosta:2003zx}
  D.~E.~Acosta {\it et al.}  [CDF II Collaboration],
  Phys.\ Rev.\ Lett.\  {\bf 93} (2004) 072001
  [arXiv:hep-ex/0312021].

\bibitem{Abazov:2004kp}
  V.~M.~Abazov {\it et al.}  [D0 Collaboration],
  Phys.\ Rev.\ Lett.\  {\bf 93} (2004) 162002
  [arXiv:hep-ex/0405004].

\bibitem{Aubert:2004ns}
  B.~Aubert {\it et al.}  [BABAR Collaboration],
  Phys.\ Rev.\  D {\bf 71} (2005) 071103
  [arXiv:hep-ex/0406022].

\bibitem{Abe:2007jn}
  K.~Abe {\it et al.}  [Belle Collaboration],
  Phys.\ Rev.\ Lett.\  {\bf 98} (2007) 082001
  [arXiv:hep-ex/0507019].

\bibitem{Abe:2007sya}
  P.~Pakhlov {\it et al.}  [Belle Collaboration],
  Phys.\ Rev.\ Lett.\  {\bf 100}, 202001 (2008)
  [arXiv:0708.3812 [hep-ex]].

\bibitem{Abe:2004zs}
  K.~Abe {\it et al.}  [Belle Collaboration],
  Phys.\ Rev.\ Lett.\  {\bf 94} (2005) 182002
  [arXiv:hep-ex/0408126].

\bibitem{Aubert:2007vj}
  B.~Aubert {\it et al.}  [BaBar Collaboration],
  Phys.\ Rev.\ Lett.\  {\bf 101} (2008) 082001
  [arXiv:0711.2047 [hep-ex]].

\bibitem{Uehara:2005qd}
  S.~Uehara {\it et al.}  [Belle Collaboration],
  Phys.\ Rev.\ Lett.\  {\bf 96}, 082003 (2006)
  [arXiv:hep-ex/0512035].


\bibitem{Barnes:2003dj}
  T.~Barnes, F.~E.~Close and H.~J.~Lipkin,
  Phys.\ Rev.\  D {\bf 68} (2003) 054006
  [arXiv:hep-ph/0305025].

\bibitem{Kolomeitsev:2003ac}
  E.~E.~Kolomeitsev and M.~F.~M.~Lutz,
  Phys.\ Lett.\  B {\bf 582} (2004) 39
  [arXiv:hep-ph/0307133].

\bibitem{Hofmann:2003je}
  J.~Hofmann and M.~F.~M.~Lutz,
  Nucl.\ Phys.\  A {\bf 733} (2004) 142
  [arXiv:hep-ph/0308263].

\bibitem{Guo:2006fu}
  F.~K.~Guo, P.~N.~Shen, H.~C.~Chiang and R.~G.~Ping,
  Phys.\ Lett.\  B {\bf 641} (2006) 278
  [arXiv:hep-ph/0603072].

\bibitem{Guo:2006rp}
  F.~K.~Guo, P.~N.~Shen and H.~C.~Chiang,
  Phys.\ Lett.\  B {\bf 647} (2007) 133
  [arXiv:hep-ph/0610008].

\bibitem{Zhang:2006ix}
  Y.~J.~Zhang, H.~C.~Chiang, P.~N.~Shen and B.~S.~Zou,
  Phys.\ Rev.\  D {\bf 74} (2006) 014013
  [arXiv:hep-ph/0604271].

\bibitem{Close:2003sg}
  F.~E.~Close and P.~R.~Page,
  Phys.\ Lett.\  B {\bf 578} (2004) 119
  [arXiv:hep-ph/0309253].

\bibitem{Wong:2003xk}
  C.~Y.~Wong,
  Phys.\ Rev.\  C {\bf 69} (2004) 055202
  [arXiv:hep-ph/0311088].

\bibitem{Swanson:2003tb}
  E.~S.~Swanson,
  Phys.\ Lett.\  B {\bf 588} (2004) 189
  [arXiv:hep-ph/0311229].

\bibitem{Gamermann:2006nm}
  D.~Gamermann, E.~Oset, D.~Strottman and M.~J.~Vicente Vacas,
  Phys.\ Rev.\  D {\bf 76} (2007) 074016
  [arXiv:hep-ph/0612179].

\bibitem{Gamermann:2007fi}
  D.~Gamermann and E.~Oset,
  Eur.\ Phys.\ J.\  A {\bf 33} (2007) 119
  [arXiv:0704.2314 [hep-ph]].

\bibitem{Gamermann:2009fv}
  D.~Gamermann and E.~Oset,
  Phys.\ Rev.\  D {\bf 80}, 014003 (2009)
  [arXiv:0905.0402 [hep-ph]].

\bibitem{Chen:2004dy}
  Y.~Q.~Chen and X.~Q.~Li,
  Phys.\ Rev.\ Lett.\  {\bf 93} (2004) 232001
  [arXiv:hep-ph/0407062].

\bibitem{Maiani:2004vq}
  L.~Maiani, F.~Piccinini, A.~D.~Polosa and V.~Riquer,
  Phys.\ Rev.\  D {\bf 71} (2005) 014028
  [arXiv:hep-ph/0412098].

\bibitem{Bracco:2005kt}
  M.~E.~Bracco, A.~Lozea, R.~D.~Matheus, F.~S.~Navarra and M.~Nielsen,
  Phys.\ Lett.\  B {\bf 624}, 217 (2005)
  [arXiv:hep-ph/0503137].

\bibitem{Bicudo:2005de}
  P.~Bicudo,
  Phys.\ Rev.\  D {\bf 74} (2006) 036008
  [arXiv:hep-ph/0512041].

\bibitem{Li:2004sta}
  B.~A.~Li,
  Phys.\ Lett.\  B {\bf 605} (2005) 306
  [arXiv:hep-ph/0410264].


\bibitem{2880-Artuso:2000xy}
  M.~Artuso {\it et al.}  [CLEO Collaboration],
  Phys.\ Rev.\ Lett.\  {\bf 86}, 4479 (2001).

\bibitem{Mizuk:2004yu}
  R.~Mizuk {\it et al.}  [Belle Collaboration],
  Phys.\ Rev.\ Lett.\  {\bf 94}, 122002 (2005).

\bibitem{Jessop:1998wt}
  C.~P.~Jessop {\it et al.}  [CLEO Collaboration],
  Phys.\ Rev.\ Lett.\  {\bf 82}, 492 (1999).

\bibitem{Csorna:2000hw}
  S.~E.~Csorna {\it et al.}  [CLEO Collaboration],
  Phys.\ Rev.\ Lett.\  {\bf 86}, 4243 (2001).

\bibitem{Chistov:2006zj}
  R.~Chistov {\it et al.}  [BELLE Collaboration],
  Phys.\ Rev.\ Lett.\  {\bf 97}, 162001 (2006).

\bibitem{Aubert:2007dt}
  B.~Aubert {\it et al.}  [BaBar Collaboration],
  Phys.\ Rev.\  D {\bf 77}, 012002 (2008).


\bibitem{Aubert:2006je}
  B.~Aubert {\it et al.}  [BaBar Collaboration],
  Phys.\ Rev.\ Lett.\  {\bf 97}, 232001 (2006).


\bibitem{2940-Aubert:2006sp}
  B.~Aubert {\it et al.}  [BaBar Collaboration],
  Phys.\ Rev.\ Lett.\  {\bf 98}, 012001 (2007).

\bibitem{2880-Abe:2006rz}
  R.~Mizuk {\it et al.}  [Belle Collaboration],
  Phys.\ Rev.\ Lett.\  {\bf 98}, 262001 (2007).


\bibitem{Tolos:2004yg}
  L.~Tolos, J.~Schaffner-Bielich and A.~Mishra,
  Phys.\ Rev.\  C {\bf 70}, 025203 (2004).

\bibitem{Lutz:2003jw}
  M.~F.~M.~Lutz and E.~E.~Kolomeitsev,
  Nucl.\ Phys.\  A {\bf 730}, 110 (2004).

\bibitem{lutz5} M.~F.~M.~Lutz and E.~E.~Kolomeitsev, Nucl.\ Phys.\ A {\bf 755}, 29c (2005).


\bibitem{Hofmann:2005sw}
  J.~Hofmann and M.~F.~M.~Lutz,
  Nucl.\ Phys.\  A {\bf 763}, 90 (2005).

\bibitem{Hofmann:2006qx}
  J.~Hofmann and M.~F.~M.~Lutz,
  Nucl.\ Phys.\  A {\bf 776}, 17 (2006).


\bibitem{Mizutani:2006vq}
  T.~Mizutani and A.~Ramos,
  Phys.\ Rev.\  C {\bf 74}, 065201 (2006).


\bibitem{GarciaRecio:2008dp}
  C.~Garcia-Recio, V.~K.~Magas, T.~Mizutani, J.~Nieves, A.~Ramos, L.~L.~Salcedo and L.~Tolos,
  Phys.\ Rev.\ D {\bf 79}, 054004 (2009).

\bibitem{JimenezTejero:2009vq}
  C.~E.~Jimenez-Tejero, A.~Ramos and I.~Vidana,
  Phys.\ Rev.\  C {\bf 80} (2009) 055206
  [arXiv:0907.5316 [hep-ph]].
  
\bibitem{meunegc}
  D.~Gamermann, C.~Garcia-Recio, J.~Nieves, L.~L.~Salcedo and L.~Tolos,
  Phys.\ Rev.\  D {\bf 81}, 094016 (2010)
  [arXiv:1002.2763 [hep-ph]].


\bibitem{Cheng:1992xi}
  H.~Y.~Cheng, C.~Y.~Cheung, G.~L.~Lin, Y.~C.~Lin, T.~M.~Yan and H.~L.~Yu,
  Phys.\ Rev.\  D {\bf 47} (1993) 1030
  [arXiv:hep-ph/9209262].

\bibitem{Cheng:1997rp}
  H.~Y.~Cheng,
  Phys.\ Lett.\  B {\bf 399} (1997) 281
  [arXiv:hep-ph/9701234].

\bibitem{Savage:1994wa}
  M.~J.~Savage,
  Phys.\ Lett.\  B {\bf 345} (1995) 61
  [arXiv:hep-ph/9408294].

\bibitem{Banuls:1999br}
  M.~C.~Banuls, A.~Pich and I.~Scimemi,
  Phys.\ Rev.\  D {\bf 61} (2000) 094009
  [arXiv:hep-ph/9911502].

\bibitem{Zhu:1998ih}
  S.~L.~F.~Zhu and Y.~B.~F.~Dai,
  Phys.\ Rev.\  D {\bf 59} (1999) 114015
  [arXiv:hep-ph/9810243].

\bibitem{Tawfiq:1999cf}
  S.~Tawfiq, J.~G.~Korner and P.~J.~O'Donnell,
  Phys.\ Rev.\  D {\bf 63} (2001) 034005
  [arXiv:hep-ph/9909444].

\bibitem{Ivanov:1999bk}
  M.~A.~Ivanov, J.~G.~Korner, V.~E.~Lyubovitskij and A.~G.~Rusetsky,
  Phys.\ Rev.\  D {\bf 60} (1999) 094002
  [arXiv:hep-ph/9904421].

\bibitem{Ivanov:1998wj}
  M.~A.~Ivanov, J.~G.~Korner and V.~E.~Lyubovitskij,
  Phys.\ Lett.\  B {\bf 448} (1999) 143
  [arXiv:hep-ph/9811370].

\bibitem{Dey:1994qi}
  J.~Dey, V.~Shevchenko, P.~Volkovitsky and M.~Dey,
  Phys.\ Lett.\  B {\bf 337} (1994) 185.

\bibitem{Fayyazuddin:1997jn}
  Fayyazuddin and Riazuddin,
  Mod.\ Phys.\ Lett.\  A {\bf 12} (1997) 1791.

\bibitem{Majethiya:2009vx}
  A.~Majethiya, B.~Patel and P.~C.~Vinodkumar,
  Eur.\ Phys.\ J.\  A {\bf 42} (2009) 213
  [arXiv:0902.2536 [hep-ph]].

\bibitem{Cho:1994vg}
  P.~L.~Cho,
  Phys.\ Rev.\  D {\bf 50} (1994) 3295
  [arXiv:hep-ph/9401276].

\bibitem{Zhu:2000py}
  S.~L.~Zhu,
  Phys.\ Rev.\  D {\bf 61} (2000) 114019
  [arXiv:hep-ph/0002023].

\bibitem{Chow:1995nw}
  C.~K.~Chow,
  Phys.\ Rev.\  D {\bf 54} (1996) 3374
  [arXiv:hep-ph/9510421].


\bibitem{Dong:2010xv}
  Y.~Dong, A.~Faessler, T.~Gutsche, S.~Kumano and V.~E.~Lyubovitskij,
  Phys.\ Rev.\  D {\bf 82}, 034035 (2010)
  [arXiv:1006.4018 [hep-ph]].
  

\bibitem{meurad}
  D.~Gamermann, L.~R.~Dai and E.~Oset,
  Phys.\ Rev.\  C {\bf 76}, 055205 (2007)
  [arXiv:0709.2339 [hep-ph]].

\bibitem{fasrad}
  A.~Faessler, T.~Gutsche, V.~E.~Lyubovitskij and Y.~L.~Ma,
  Phys.\ Rev.\  D {\bf 76}, 014005 (2007)
  [arXiv:0705.0254 [hep-ph]].

\bibitem{lutz1}
  M.~F.~M.~Lutz and M.~Soyeur,
  Nucl.\ Phys.\  A {\bf 813} (2008) 14
  [arXiv:0710.1545 [hep-ph]].

\bibitem{yubid}
  Y.~Dong, A.~Faessler, T.~Gutsche and V.~E.~Lyubovitskij,
  arXiv:0909.0380 [hep-ph].

\bibitem{molina1}
  W.~H.~Liang, R.~Molina and E.~Oset,
  Eur.\ Phys.\ J.\  A {\bf 44}, 479 (2010)
  [arXiv:0912.4359 [hep-ph]].

\bibitem{molina2}
  R.~Molina, H.~Nagahiro, A.~Hosaka and E.~Oset,
  arXiv:1009.4881 [hep-ph].

\bibitem{molina3}
  T.~Branz, R.~Molina and E.~Oset,
  arXiv:1010.0587 [hep-ph].


\bibitem{pdg}
  C.~Amsler {\it et al.}  [Particle Data Group],
  Phys.\ Lett.\  B {\bf 667}, 1 (2008).

\bibitem{machleidt}
  R.~Machleidt, K.~Holinde and C.~Elster,
  Phys.\ Rept.\  {\bf 149}, 1 (1987).

\bibitem{Logan:1967zz}
  R.~K.~Logan and H.~W.~Wyld,
  Phys.\ Rev.\  {\bf 158}, 1467 (1967).

\bibitem{meisner}
  B.~Borasoy, P.~C.~Bruns, U.~G.~Meissner and R.~Nissler,
  Phys.\ Rev.\  C {\bf 72}, 065201 (2005)
  [arXiv:hep-ph/0508307].

\bibitem{doering1}
  M.~Doring, E.~Oset and S.~Sarkar,
  Phys.\ Rev.\  C {\bf 74}, 065204 (2006)
  [arXiv:nucl-th/0601027].

\bibitem{doering2}
  M.~Doring,
  Nucl.\ Phys.\  A {\bf 786}, 164 (2007)
  [arXiv:nucl-th/0701070].

\bibitem{geng}
  L.~S.~Geng, E.~Oset and M.~Doring,
  Eur.\ Phys.\ J.\  A {\bf 32}, 201 (2007)
  [arXiv:hep-ph/0702093].

\bibitem{ramos1}
  E.~Oset and A.~Ramos,
  Nucl.\ Phys.\  A {\bf 635}, 99 (1998)
  [arXiv:nucl-th/9711022].



\end{thebibliography}
\end{document}